\documentclass[twocolumn,aps,prl,superscriptaddress,showpacs,amsmath,amssymb]{revtex4}

\usepackage{graphicx}
\usepackage{dcolumn}
\usepackage{bm}

\begin{document}


\title{Spin-Peierls transition in TiOCl}

\author{Mohammad Shaz}
\author{Sander van Smaalen}
\email{smash@uni-bayreuth.de}
\homepage{http://www.uni-bayreuth.de/departments/crystal/}
\affiliation{Laboratory of Crystallography, University of Bayreuth,
95440 Bayreuth, Germany}
\author{Lukas Palatinus}
\affiliation{Laboratory of Crystallography, University of Bayreuth,
95440 Bayreuth, Germany}
\affiliation{Institute of Physics of the Academy of Sciences of
the Czech republic, Prag, Czech Republic}%
\author{Markus Hoinkis}
\author{Matthias Klemm}
\author{Siegfried Horn}
\author{Ralph Claessen}
\affiliation{Experimentalphysik II, University of Augsburg,
D-86135 Augsburg, Germany}%

\date{\today}

\begin{abstract}

Temperature-dependent x-ray diffraction of the low-dimensional
spin 1/2 quantum magnet TiOCl shows that the phase
transition at $T_{c2}=90$ K corresponds to a lowering of the
lattice symmetry.
Below $T_{c1}=66$ K a twofold superstructure develops,
that indicates the formation of spin-singlet
pairs via direct exchange between neighboring Ti atoms, while
the role of superexchange is found to be negligible.
TiOCl thus is identified as a spin-Peierls system of
pure 1D chains of atoms.
The first-order character of the transition at $T_{c1}$ is
explained by the competition
between the structurally deformed state below $T_{c2}$ and
the spin-Peierls state below $T_{c1}$.
\end{abstract}

\pacs{61.50.Ks, 75.30.Kz, 75.30.Et, 61.66.Fn}
\maketitle

Low-dimensional $S=1/2$ quantum spin systems are of interest,
because of the importance for understanding the mechanism
of high-$T_c$ superconductivity \cite{beynonrj1993},
as well as for their potential applications in quantum computers.
Furthermore, the relatively simple, yet complicated materials with
$1$-dimensional (1D) or 2D quantum spin systems offer a
wide variety of ground states, that are accessible by
\textit{ab initio} theory, and therefore might help towards
the understanding of fundamental quantum mechanical properties
of solids.
The development of magnetic order at low temperatures
may or may not be coupled to a change of the electronic
structure, resulting in ground states with antiferromagnetic
order or spin-density waves (SDW).
The spin-Peierls state is defined by singlet
pairs of localized electrons, that form because of
an enhancement of exchange interactions between neighboring
magnetic atoms due to a dimerization of the crystal structure.


CuGeO$_3$ is the only inorganic compound for which the
spin-Peierls state below $T_c=14$ K has been
unambiguously established
\cite{hasem1993,pouget1994,kamimura1994,hirotak1994}.
Initially NaV$_2$O$_5$ was considered to be a candidate
spin-Peierls material, but more recent work showed that
the $4$-fold superstructure below $T_c=34$ K is related
to a combination of charge-, orbital- and magnetic order
\cite{lemmensp2003b}.
Recently, Seidel \textit{et al.} \cite{seidela2003}
proposed that TiOCl is a 1D $S=1/2$ quantum spin system,
that transforms into a spin-Peierls state at low temperatures.

\begin{figure}
\begin{minipage}[b]{8.0cm}
{\includegraphics[width=7.8cm]{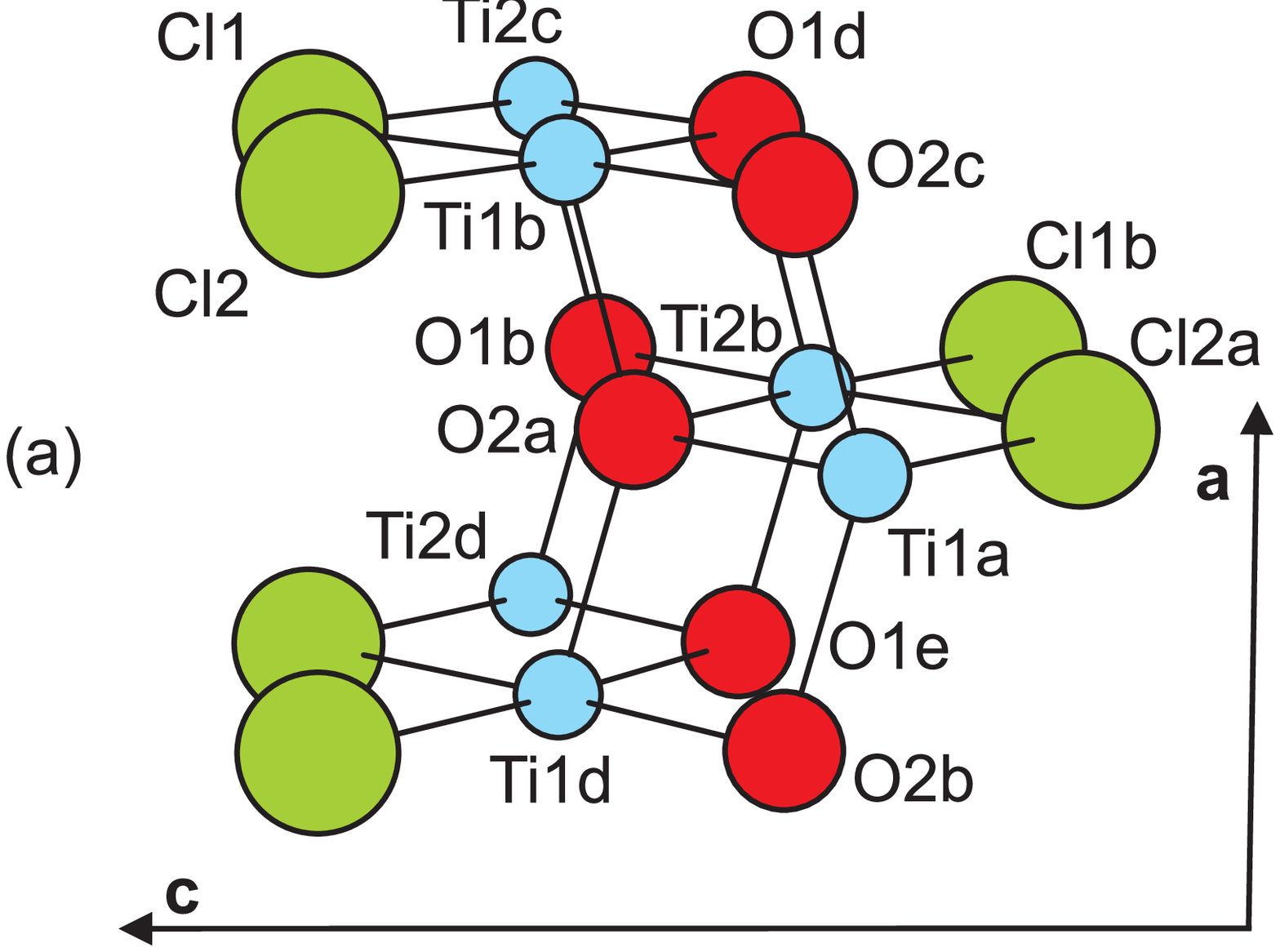}}
\end{minipage}
\begin{minipage}[b]{8.0cm}
{\includegraphics[width=7.8cm]{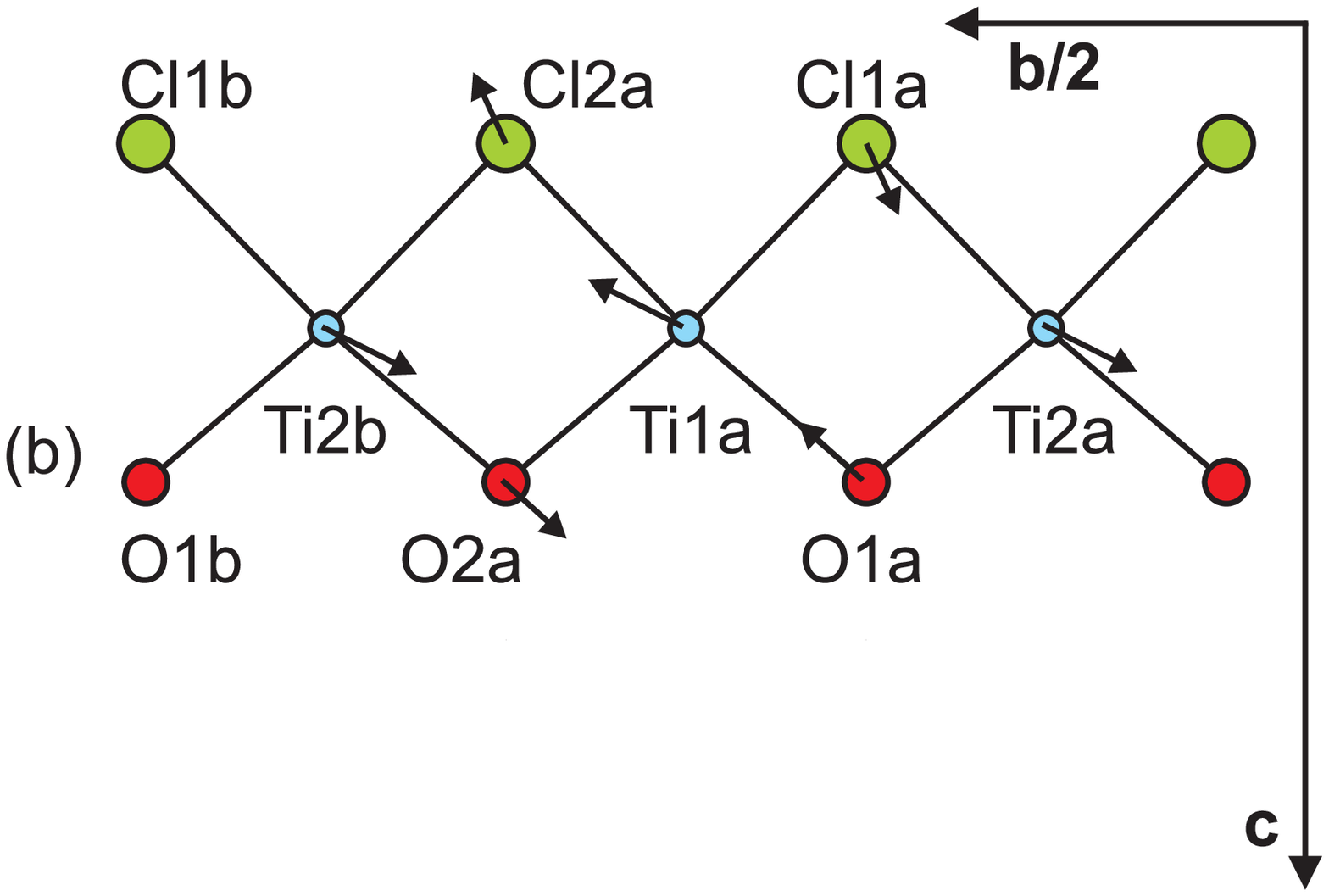}}
\end{minipage}
\caption{\label{f-basic_structure}
Crystal structure of TiOCl.
(a) Perspective view of one layer.
(b) The ribbon parallel $\mathbf{b}$ at $x=0$, containing a
chain of Ti atoms;
The displacements in the superstructure are given by arrows
($20\times$ the true values).
Crystallographically independent atoms are indicated by numbers.
Different but symmetry equivalent atoms are indicated by an
additional letter, that corresponds to Table \ref{t-distances}.
The unit cell axes are indicated.
(color online: blue for Ti, red for O, and green for Cl.)}
\end{figure}

TiOCl crystallizes in a layered structure \cite{schaferh1958}
(Fig. \ref{f-basic_structure}), in which two different types
of chains of Ti$^{3+}$, $d^1$ ($S=1/2$) have been
identified \cite{seidela2003}.
The chain along $\mathbf{a}$ allows interactions between
the electrons via superexchange, whereas the chain along $\mathbf{b}$
supports direct exchange interactions.
The latter type of chains has been proposed to be
responsible for the quasi-1-dimensional (1D) character
of the magnetic interactions, as evidenced by the magnetic
susceptibility, electron spin resonance (ESR), IR reflectivity,
angle-resolved photoelectron spectroscopy (ARPES)
and electronic band structure calculations
\cite{seidela2003,kataevv2003,caimig2004a,hoinkism2004,sahadasgupta2004}.
Based on the temperature dependencies of the
magnetic susceptibility \cite{seidela2003}, ESR \cite{kataevv2003}
and NMR \cite{imait2003} a second-order phase transition
was found at $T_{c2}=94$ K, while a first-order
transition takes place at $T_{c1}=67$ K.
The latter transition corresponds to a sudden development
of magnetic order,
accompanied by a doubling of the lattice constant along
$\mathbf{b}$ \cite{seidela2003,imait2003}.
The magnetic moments are zero below $T_{c1}$, and
the size ($E_g=430$ K) of the spin-gap has
been taken as an indication for a non-conventional
spin-Peierls state at low temperatures \cite{imait2003}.
Above $T_{c2}$ up to $T^{*}=135$ K a pseudo spin-gap
due to fluctuations has been found \cite{imait2003,caimig2004a}.

In the present contribution we report the discovery of superlattice
reflections in the x-ray diffraction of TiOCl below
$T_{c1}=66$ K as well as a complete structure
determination at $T=10$ K.
The latter shows that
the 2-fold superstructure can be interpreted as a dimerization
of the 1D chains of Ti atoms along $\mathbf{b}$, while the
interatomic distances and bond angles between chains
are much less affected by the structural deformation.
Our results thus indicate that -- despite the unconventional
magnetic behavior observed at higher temperatures -- the
opening of the spin gap below 66 K is due to a spin-Peierls
transition, where a surprisingly simple dimerization of the
1D Ti atom chains is responsible for the formation of spin
singlets.

Single crystals of TiOCl were prepared by gas transport
in evacuated quartz glass tubes, following published
procedures \cite{schaferh1958}.
Single-crystal x-ray diffraction with synchrotron radiation was performed at
beam-line D3 of Hasylab (DESY, Hamburg), employing
monochromatized radiation of wavelength 0.5000 {\AA}.
A single crystal of dimensions
$0.05 \times 0.11 \times 0.01$ mm$^{3}$
was mounted on a carbon fibre attached to a closed-cycle
helium cryostat mounted on a Huber 4-circle diffractometer.
X-ray diffraction was measured by a point detector.

\begin{figure}
{\includegraphics[width=7.8cm]{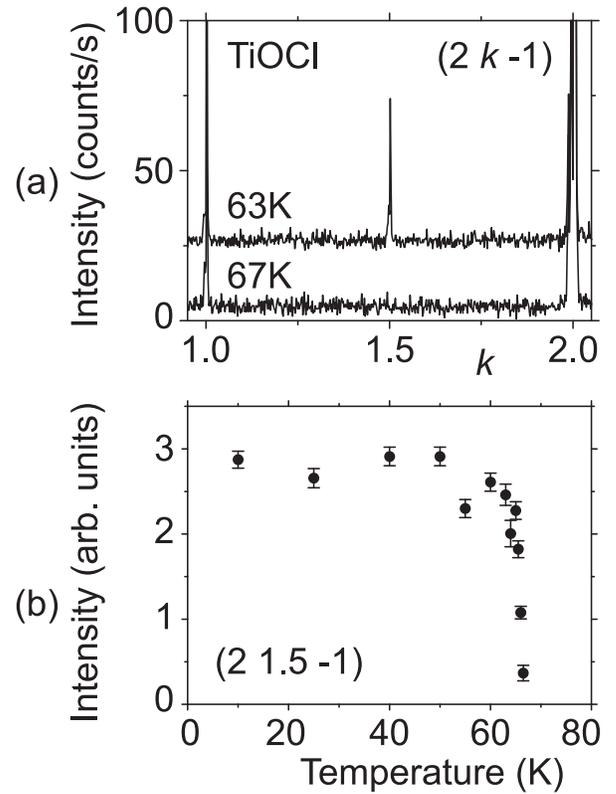}}
\caption{\label{f-superlattice_reflections}
Superlattice reflections of TiOCl below $T_{c1}=66$K.
(a) $q$-scan $(2, 1+\upsilon, -1)$ with $-0.05 < \upsilon < 1.05$ at
temperatures of 63K (shifted by 30 counts/s) and 67K.
(b) Temperature dependence of the integrated intensities
of the superlattice reflection $(2, 1.5, -1)$.
The intensity at $T=67$K was measured as less than zero
(zero within standard uncertainty).
Error bars indicate standard uncertainties.}
\end{figure}

\begin{figure}
{\includegraphics[width=7.8cm]{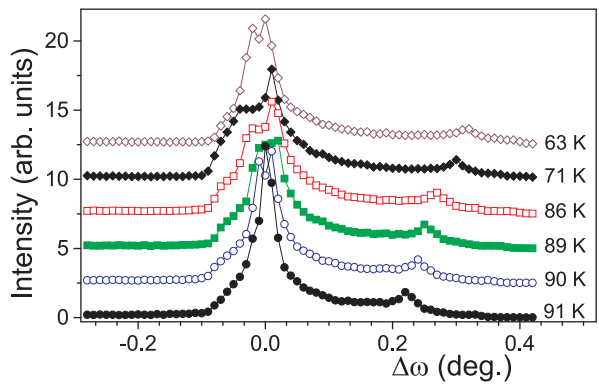}}
\caption{\label{f-reflection_width}
$\omega$-scans of the $(0, 2, 0)$ main reflection at
several temperatures.
Scans at $T>91$K (data not shown) exhibit similar widths
as the scan at 91 K.
Each scan is centered on $\Delta\,\omega=0$, and
an incremental offset of 2.5 has been applied.
The large width near the base of the reflections and
the secondary maximum are the result of the relatively
poor quality of the single crystal.
}
\end{figure}

In a first experiment $q$-scans were performed at $T=10$ K along
the reciprocal lattice lines $(h+\xi,k,l)$, $(h,k+\xi,l)$,
$(h+1/2,k+\xi,l)$, $(h+1/2,k+\xi,l+1/2)$ and
$(h,k+\xi,l+1/2)$ for approximately 20 different $(h,k,l)$
combinations, and with $\xi$ scanned from $0$ to $1$.
Superlattice reflections were only found at $(h,k+1/2,l)$ positions,
indicating a doubling of the unit cell along $\mathbf{b}$
(Fig. \ref{f-superlattice_reflections}a).
A few of the strongest superlattice reflections were selected
for temperature-dependent measurements.
$q$-scans established that the relative positions of the
superlattice reflections are independent of temperature.
The integrated intensities were measured by $\omega$-scans.
Up to about 65K, they were found to be independent
of temperature, but then they dropped continuously to
zero at 67K (Fig. \ref{f-superlattice_reflections}b),
which can be explained
by the coexistence of the intermediate- and low-temperature phases,
as can be the result of internal strains within the sample.
$q$-scans above 67 K did not give any evidence
for an incommensurate superstructure.
Scans of several main reflections showed that for
$T\le 90$ K their full width at half maximum (FWHM)
is larger than for $T\ge 91$ K
(Fig. \ref{f-reflection_width}).
The broadening was reversible and gradually increased
on decreasing temperature, but it did not show a clear
anomaly at $T_{c1}$.
The splitting of reflections was not resolved, thus
preventing the determination of the space group
of the intermediate phase.

These results indicate a second-order phase transition
at $T_{c2}=90 (1)$ K corresponding to a lowering of the
point symmetry of the crystal structure,
followed by a first-order phase transition
at $T_{c1} = 66 (1)$K towards a structure with a doubled
$\mathbf{b}$ axis.
The $q$-scans at 67K show that the transition at $T_{c1}$
does not correspond to an incommensuration of the superstructure,
contrary to the suggestion by NMR \cite{imait2003}.
Alternatively, the reduced lattice symmetry below $T_{c2}$
offers an explanation for the broad resonances in NMR.

At $T=10$K we have measured the integrated intensities
of all Bragg reflections up to $(\sin(\theta)/\lambda)_{max}$
= 0.7 {\AA}$^{-1}$.
Structure refinements were performed within the superspace
approach applied to this commensurately modulated structure
\cite{dewolff1981,vansmaalen1995},
with the superspace group P$mmn(0,0.5,0)$
and with lattice parameters $a=3.7829$, $b_0=3.3415$
and $c=8.0305$ {\AA} at $T=10$K.
Different sections $t_0$ of superspace correspond to
superstructures $a\times 2b_0\times c$ with
different symmetries \cite{vansmaalen1995}.
Under the assumption of twinning, the best fit to the
diffraction data was obtained for $t_0=0.125$, corresponding
to a supercell with monoclinic symmetry and
space group P$2_1/m$ ($\mathbf{a}$ axis unique).
The fit with $t_0 =0.0$ was nearly as good.
It corresponds to a supercell with acentric orthorhombic
symmetry P$mm2$
\footnote{758 measured reflections reduce to 252 observed unique
main reflections
and 153 observed unique superlattice reflections in monoclinic symmetry.
The refinement converged to a model with reliability factors
$R_{all}=0.055$ with $R_{main}=0.051$ and $R_{sat}=0.083$.
In orthorhombic symmetry 225 unique main reflections and
130 unique satellites were obtained.
The refinement converged at $R_{all}=0.058$
with $R_{main}=0.055$ and $R_{sat}=0.081$.}.
The two structure models are completely different.
P$mm2$ involves three crystallographically independent
Ti and Cl atoms, that are distributed over two independent
ribbons along $\mathbf{b}$.
P$2_1/m$ involves two independent Ti and Cl atoms, that
alternate along a single unique ribbon along $\mathbf{b}$
(Fig. \ref{f-basic_structure}).
NMR indicates two crystallographically independent Ti and Cl
atoms at low temperatures \cite{imait2003}.
It can therefore be concluded, that the superlattice structure
has monoclinic symmetry P$2_1/m$.



One ribbon parallel $\mathbf{b}$ contains the 6 crystallographically
independent atoms in the supercell P$2_1/m$, all of which are
in the mirror plane.
Accordingly, the displacements of the atoms are restricted to
the $\mathbf{b},\mathbf{c}$ plane (Fig. \ref{f-basic_structure}b).
The pattern of displacements clearly shows a dimerization
of the chain of Ti atoms, while the displacements of
O and Cl atoms are such as to minimize the elastic strain
of the structure.
This interpretation is supported by comparing the interatomic
distances in the superstructure (Table \ref{t-distances}) and
the basic structure (Table \ref{t-basic_geometry10+295k}).
The largest variation of 0.18 {\AA} is found
for the alternating distances on the Ti chains
along $\mathbf{b}$, indicating the formation of
Ti--Ti pairs on these chains.
The shortest Ti--Ti distance is between chains,
but its variation in the superstructure is much less than
the variation of distances along the chains, thus supporting
the model of pair building on the chains as opposed
to the formation of singlet pairs between electrons
on neighboring chains.
The smallest variation (0.024 {\AA}) is obtained for
the four independent Ti--Cl distances,
while the variation of Ti--O distances (0.049 {\AA}) is
still much smaller than the
variation of Ti--Ti distances along the chains.
This interpretation of the superstructure supports the model
of spin pairing on chains of Ti atoms along $\mathbf{b}$
via direct exchange \cite{seidela2003}.
Although bond angles Ti--O--Ti and Ti--Cl--Ti
alternate along the chains, they do not provide evidence
for a role of superexchange in the spin-Peierls transition.

\begin{table}
\caption{\label{t-distances}
Selected interatomic distances and bond angles in the superstructure of TiOCl at 10K.
Crystallographically independent atoms are indicated by numbers
1 and 2.
Different but symmetry equivalent atoms are indicated by an
additional letter.
Standard uncertainties are given in parentheses.}
\begin{ruledtabular}
\begin{tabular}{l l @{\hspace{0.2cm}} l l}
Atoms   &  Distance ({\AA}) & Atoms & Distance ({\AA}) \\
\hline
Ti1a--Ti2a\footnotemark[1]& 3.429 (1)&
Ti1a--Ti1b\footnotemark[2]& 3.159 (1) \\
Ti1a--Ti2b\footnotemark[1]& 3.254 (1)&
Ti1b--Ti2b\footnotemark[2]& 3.177 (1) \\
Ti1b--Ti1d\footnotemark[3]& 3.783 &
Ti2b--Ti2c\footnotemark[2]& 3.198 (1) \\
Ti2c--Ti2d\footnotemark[3]& 3.783 \\
Ti1a--O1a\footnotemark[1]   & 2.203 (3)& Ti2b--O1b\footnotemark[1] & 2.222 (3)\\
Ti1a--O2a\footnotemark[1]  & 2.173 (4)& Ti2b--O2a\footnotemark[1]  & 2.189 (3)\\
Ti1a--O2b\footnotemark[2]  & 1.956 (1)& Ti2b--O1d\footnotemark[2] & 1.959 (1)\\
Ti1a--O2c\footnotemark[2]  & 1.956 (1)& Ti2b--O1e\footnotemark[2] & 1.959 (1)\\
Ti1a--Cl1a\footnotemark[1]  & 2.398 (2)& Ti2b--Cl1b\footnotemark[1] & 2.407 (2)\\
Ti1a--Cl2a\footnotemark[1]  & 2.383 (2)& Ti2b--Cl2a\footnotemark[1] & 2.394 (2)\\
\hline
Atoms   &  Angle ({deg.}) & Atoms & Angle (deg.) \\
\hline
Ti1a--O1a--Ti2a\footnotemark[1] & 101.6 (2)
             & Ti1a--Cl1a--Ti2a\footnotemark[1] & 91.1 (1) \\
Ti1a--O2a--Ti2b\footnotemark[1] & 96.5 (2)
             & Ti1a--Cl2a--Ti2b\footnotemark[1] & 85.9 (1) \\
Ti1a--O2a--Ti1b\footnotemark[2] & 99.7 (1) &
            Ti1b--O1d--Ti2b\footnotemark[2] & 99.4 (1) \\
Ti2b--O1b--Ti2c\footnotemark[2] & 99.6 (1) &
            Ti1b--O2a--Ti2b\footnotemark[2] & 99.9 (1) \\
Ti1b--O2a--Ti1d\footnotemark[3] & 150.4 (2) &
            Ti2c--O1b--Ti2d\footnotemark[3] & 149.8 (2) \\
\end{tabular}
\end{ruledtabular}
\footnotetext[1]{Distances and angles within a single ribbon parallel $\mathbf{b}$.}
\footnotetext[2]{Distances and angles between neighboring ribbons, $\mathbf{a}/2$ apart.}
\footnotetext[3]{Distances and angles towards ribbons at $\pm\mathbf{a}$.}
\end{table}

\begin{table}
\caption{\label{t-basic_geometry10+295k}
Distances ({\AA}) and bond angles (deg.) in the basic structures
of TiOCl at room temperature and at $T=10$K.
One crystallographically independent atom each exists for Ti, O and Cl.
}
\begin{ruledtabular}
\begin{tabular}{l l l}
Atoms                       & $T=10$K   & $T=295$K \\
\hline
Ti--Ti\footnotemark[1]      & 3.342     & 3.355 \\
Ti--O\footnotemark[1]       & 2.196 (1) & 2.187 (2) \\
Ti--Cl\footnotemark[1]      & 2.395 (1) & 2.393 (1) \\
Ti--O--Ti\footnotemark[1]   & 99.2 (1)  & 100.2 (1) \\
Ti--Cl--Ti\footnotemark[1]  & 88.4 (1)  & 89.0 (1) \\
Ti--Ti\footnotemark[2]      & 3.177 (1)  & 3.172 (1) \\
Ti--O\footnotemark[2]       & 1.958 (1) & 1.958 (1) \\
Ti--O--Ti\footnotemark[2]   & 99.7 (1)   & 99.7 (1) \\
Ti--Ti\footnotemark[3]      & 3.783      & 3.779 \\
Ti--O--Ti\footnotemark[3]   & 150.0 (1)  & 149.5 (2) \\
\end{tabular}
\end{ruledtabular}
\footnotetext[1]{Distances and angles within a single ribbon parallel $\mathbf{b}$.}
\footnotetext[2]{Distances and angles between neighboring ribbons, $\mathbf{a}/2$ apart.}
\footnotetext[3]{Distances and angles towards ribbons at $\pm\mathbf{a}$.}
\end{table}

The crystal structure at room temperature and the
basic structure at 10 K are nearly equal, with
the small differences in interatomic distances
explained by thermal expansion (Table \ref{t-basic_geometry10+295k}).
This implies that structural changes within P$mmn$ cannot
be at the origin of the observed anomalies at $T^{*}$.
Instead
the anomalous behavior in the temperature
dependencies of the magnetic susceptibility and
spectroscopic properties below $T^* = 135$ K \cite{imait2003}
might be related to a temperature dependence of
the fluctuations.
Furthermore we did not find any evidence for a possible
lower symmetry than P$mmn$ at room temperature,
as it was proposed by
Caimi \textit{et al.} \cite{caimig2004b,caimig2004a}.
We failed to observe incommensurate superlattice reflections for
the phase between $T_{c2}$ and $T_{c1}$, as
was proposed in \cite{imait2003}.
Instead our results suggest a significant deviation of the
structure from P$mmn$ between $T_{c1}$ and $T_{c2}$,
that is replaced by a 2-fold superstructure below $T_{c1}$,
while an average P$mmn$ symmetry is restored for the
basic structure.
This scenario offers an explanation for the first-order
character of the transition at $T_{c1}$.
The envisaged second-order spin-Peierls transition from
P$mmn$ towards the observed superstructure would have occurred
at $T_{SP}<T_{c2}$, but it is superseded
by a structural transition at $T_{c2}$ towards the
intermediate phase
with a distortion that is different from the dimerization
in the spin-Peierls state.
Below $T_{c1}<T_{SP}$
the spin-Peierls state becomes the most stable state.
It develops out of the intermediate phase, thus requiring
major structural rearrangements, which explains the first-order
character of the phase transition at $T_{c1}$.
Band-structure calculations have found that the
electronic ground state for the room-temperature structure
is characterized by a  $d_{xy}$$^{1}$ configuration of the
Ti$^{3+}$ ion \cite{sahadasgupta2004}.
Accordingly it should also be the ground state at low temperatures.
The transitions at $T_{c1}$ then would lead to a further
stabilization of the $d_{xy}$ orbital, and the suppression
of fluctuations involving other $d$ orbitals, but it would
not correspond to orbital order.


In conclusion, we have found that TiOCl exhibits a first-order
phase transition at $T_{c1}=66 (1)$ K towards a superstructure
with a doubled lattice constant $\mathbf{b}$ along the chains of
Ti atoms and with monoclinic symmetry P$2_1/m$,
while at $T_{c2}=90 (1)$ K the lattice symmetry
has already become lower than orthorhombic.
The superstructure is characterized by a simple
dimerization of the chains of Ti atoms, while the
displacements of the O and Cl atoms obtain values
to minimize the internal strain of the superstructure.
Despite the unconventional magnetic behavior observed
at higher temperatures,
our results provide compelling evidence for
a spin-Peierls state of TiOCl below $T_{c1} = 66 (1)$K,
that is achieved through direct exchange between
electron spins on the chains of Ti atoms parallel
$\mathbf{b}$ \cite{seidela2003}.
The superstructure of TiOCl is surprisingly simple,
much simpler than that of
CuGeO$_3$ \cite{hirotak1994}, and TiOCl can thus be identified as
a spin-Peierls system with the relevant
magnetic interactions confined to a true 1D chain of atoms,
and with the highest transition temperature yet observed.

\begin{acknowledgments}
We gratefully acknowledge support with
the synchrotron experiment by W. Morgenroth.
X-ray diffraction with synchrotron radiation was performed
at beam-line D3 of Hasylab at DESY (Hamburg, Germany).
This work was supported by the German Science Foundation
(DFG) through grants SM55/6-3, CL124/3-3, and SFB484.
\end{acknowledgments}


\begin{thebibliography}{16}
\expandafter\ifx\csname natexlab\endcsname\relax\def\natexlab#1{#1}\fi
\expandafter\ifx\csname bibnamefont\endcsname\relax
  \def\bibnamefont#1{#1}\fi
\expandafter\ifx\csname bibfnamefont\endcsname\relax
  \def\bibfnamefont#1{#1}\fi
\expandafter\ifx\csname citenamefont\endcsname\relax
  \def\citenamefont#1{#1}\fi
\expandafter\ifx\csname url\endcsname\relax
  \def\url#1{\texttt{#1}}\fi
\expandafter\ifx\csname urlprefix\endcsname\relax\def\urlprefix{URL }\fi
\providecommand{\bibinfo}[2]{#2}
\providecommand{\eprint}[2][]{\url{#2}}

\bibitem[{\citenamefont{Beynon and Wilson}(1993)}]{beynonrj1993}
\bibinfo{author}{\bibfnamefont{R.~J.} \bibnamefont{Beynon}} \bibnamefont{and}
  \bibinfo{author}{\bibfnamefont{J.~A.} \bibnamefont{Wilson}},
  \bibinfo{journal}{J. Phys.: Condens. Matter} \textbf{\bibinfo{volume}{5}},
  \bibinfo{pages}{1983} (\bibinfo{year}{1993}).

\bibitem[{\citenamefont{Hase et~al.}(1993)\citenamefont{Hase, Terasaki, and
  Uchinokura}}]{hasem1993}
\bibinfo{author}{\bibfnamefont{M.}~\bibnamefont{Hase}},
  \bibinfo{author}{\bibfnamefont{I.}~\bibnamefont{Terasaki}}, \bibnamefont{and}
  \bibinfo{author}{\bibfnamefont{K.}~\bibnamefont{Uchinokura}},
  \bibinfo{journal}{Phys. Rev. Lett.} \textbf{\bibinfo{volume}{70}},
  \bibinfo{pages}{3651} (\bibinfo{year}{1993}).

\bibitem[{\citenamefont{Pouget et~al.}(1994)\citenamefont{Pouget, Regnault,
  Ain, Hennion, Renard, Veillet, Dhalenne, and Revcolevschi}}]{pouget1994}
\bibinfo{author}{\bibfnamefont{J.~P.} \bibnamefont{Pouget}},
  \bibinfo{author}{\bibfnamefont{L.~P.} \bibnamefont{Regnault}},
  \bibinfo{author}{\bibfnamefont{M.}~\bibnamefont{Ain}},
  \bibinfo{author}{\bibfnamefont{B.}~\bibnamefont{Hennion}},
  \bibinfo{author}{\bibfnamefont{J.~P.} \bibnamefont{Renard}},
  \bibinfo{author}{\bibfnamefont{P.}~\bibnamefont{Veillet}},
  \bibinfo{author}{\bibfnamefont{G.}~\bibnamefont{Dhalenne}}, \bibnamefont{and}
  \bibinfo{author}{\bibfnamefont{A.}~\bibnamefont{Revcolevschi}},
  \bibinfo{journal}{Phys. Rev. Lett.} \textbf{\bibinfo{volume}{72}},
  \bibinfo{pages}{4037} (\bibinfo{year}{1994}).

\bibitem[{\citenamefont{Kamimura et~al.}(1994)\citenamefont{Kamimura, Terauchi,
  Tanaka, Fujita, and Akimitsu}}]{kamimura1994}
\bibinfo{author}{\bibfnamefont{O.}~\bibnamefont{Kamimura}},
  \bibinfo{author}{\bibfnamefont{M.}~\bibnamefont{Terauchi}},
  \bibinfo{author}{\bibfnamefont{M.}~\bibnamefont{Tanaka}},
  \bibinfo{author}{\bibfnamefont{O.}~\bibnamefont{Fujita}}, \bibnamefont{and}
  \bibinfo{author}{\bibfnamefont{J.}~\bibnamefont{Akimitsu}},
  \bibinfo{journal}{J. Phys. Soc. Jpn.} \textbf{\bibinfo{volume}{63}},
  \bibinfo{pages}{2467} (\bibinfo{year}{1994}).

\bibitem[{\citenamefont{Hirota et~al.}(1994)\citenamefont{Hirota, Cox, Lorenzo,
  Shirane, Tranquada, Hase, Uchinokura, Kojima, Shibuya, and
  Tanaka}}]{hirotak1994}
\bibinfo{author}{\bibfnamefont{K.}~\bibnamefont{Hirota}},
  \bibinfo{author}{\bibfnamefont{D.~E.} \bibnamefont{Cox}},
  \bibinfo{author}{\bibfnamefont{J.~E.} \bibnamefont{Lorenzo}},
  \bibinfo{author}{\bibfnamefont{G.}~\bibnamefont{Shirane}},
  \bibinfo{author}{\bibfnamefont{J.~M.} \bibnamefont{Tranquada}},
  \bibinfo{author}{\bibfnamefont{M.}~\bibnamefont{Hase}},
  \bibinfo{author}{\bibfnamefont{K.}~\bibnamefont{Uchinokura}},
  \bibinfo{author}{\bibfnamefont{H.}~\bibnamefont{Kojima}},
  \bibinfo{author}{\bibfnamefont{Y.}~\bibnamefont{Shibuya}}, \bibnamefont{and}
  \bibinfo{author}{\bibfnamefont{I.}~\bibnamefont{Tanaka}},
  \bibinfo{journal}{Phys. Rev. Lett.} \textbf{\bibinfo{volume}{73}},
  \bibinfo{pages}{736} (\bibinfo{year}{1994}).

\bibitem[{\citenamefont{Lemmens et~al.}(2003)\citenamefont{Lemmens,
  G\"untherodt, and Gros}}]{lemmensp2003b}
\bibinfo{author}{\bibfnamefont{P.}~\bibnamefont{Lemmens}},
  \bibinfo{author}{\bibfnamefont{G.}~\bibnamefont{G\"untherodt}},
  \bibnamefont{and} \bibinfo{author}{\bibfnamefont{C.}~\bibnamefont{Gros}},
  \bibinfo{journal}{Phys. Rep.} \textbf{\bibinfo{volume}{375}},
  \bibinfo{pages}{1} (\bibinfo{year}{2003}).

\bibitem[{\citenamefont{Seidel et~al.}(2003)\citenamefont{Seidel, Marianetti,
  Chou, Ceder, and Lee}}]{seidela2003}
\bibinfo{author}{\bibfnamefont{A.}~\bibnamefont{Seidel}},
  \bibinfo{author}{\bibfnamefont{C.~A.} \bibnamefont{Marianetti}},
  \bibinfo{author}{\bibfnamefont{F.~C.} \bibnamefont{Chou}},
  \bibinfo{author}{\bibfnamefont{G.}~\bibnamefont{Ceder}}, \bibnamefont{and}
  \bibinfo{author}{\bibfnamefont{P.~A.} \bibnamefont{Lee}},
  \bibinfo{journal}{Phys. Rev. B} \textbf{\bibinfo{volume}{67}},
  \bibinfo{pages}{020405} (\bibinfo{year}{2003}).

\bibitem[{\citenamefont{Sch\"afer et~al.}(1958)\citenamefont{Sch\"afer,
  Wartenpfuhl, and Weise}}]{schaferh1958}
\bibinfo{author}{\bibfnamefont{H.}~\bibnamefont{Sch\"afer}},
  \bibinfo{author}{\bibfnamefont{F.}~\bibnamefont{Wartenpfuhl}},
  \bibnamefont{and} \bibinfo{author}{\bibfnamefont{E.}~\bibnamefont{Weise}},
  \bibinfo{journal}{Z. Anorg. Allg. Chemie} \textbf{\bibinfo{volume}{295}},
  \bibinfo{pages}{268} (\bibinfo{year}{1958}).

\bibitem[{\citenamefont{Kataev et~al.}(2003)\citenamefont{Kataev, Baier,
  Moller, Jongen, Meyer, and Freimuth}}]{kataevv2003}
\bibinfo{author}{\bibfnamefont{V.}~\bibnamefont{Kataev}},
  \bibinfo{author}{\bibfnamefont{J.}~\bibnamefont{Baier}},
  \bibinfo{author}{\bibfnamefont{A.}~\bibnamefont{Moller}},
  \bibinfo{author}{\bibfnamefont{L.}~\bibnamefont{Jongen}},
  \bibinfo{author}{\bibfnamefont{G.}~\bibnamefont{Meyer}}, \bibnamefont{and}
  \bibinfo{author}{\bibfnamefont{A.}~\bibnamefont{Freimuth}},
  \bibinfo{journal}{Phys. Rev. B} \textbf{\bibinfo{volume}{68}},
  \bibinfo{pages}{140405} (\bibinfo{year}{2003}).

\bibitem[{\citenamefont{Caimi et~al.}(2004{\natexlab{a}})\citenamefont{Caimi,
  Degiorgi, Kovaleva, Lemmens, and Chou}}]{caimig2004a}
\bibinfo{author}{\bibfnamefont{G.}~\bibnamefont{Caimi}},
  \bibinfo{author}{\bibfnamefont{L.}~\bibnamefont{Degiorgi}},
  \bibinfo{author}{\bibfnamefont{N.~N.} \bibnamefont{Kovaleva}},
  \bibinfo{author}{\bibfnamefont{P.}~\bibnamefont{Lemmens}}, \bibnamefont{and}
  \bibinfo{author}{\bibfnamefont{F.~C.} \bibnamefont{Chou}},
  \bibinfo{journal}{Phys. Rev. B} \textbf{\bibinfo{volume}{69}},
  \bibinfo{pages}{125108} (\bibinfo{year}{2004}{\natexlab{a}}).

\bibitem[{\citenamefont{Hoinkis and \textit{et al.}}(2004)}]{hoinkism2004}
\bibinfo{author}{\bibfnamefont{M.}~\bibnamefont{Hoinkis}} \bibnamefont{and}
  \bibinfo{author}{\bibnamefont{\textit{et al.}}},
  \bibinfo{journal}{unpublished}  (\bibinfo{year}{2004}).

\bibitem[{\citenamefont{Saha-Dasgupta et~al.}(2004)\citenamefont{Saha-Dasgupta,
  Valenti, Rosner, and Gros}}]{sahadasgupta2004}
\bibinfo{author}{\bibfnamefont{T.}~\bibnamefont{Saha-Dasgupta}},
  \bibinfo{author}{\bibfnamefont{R.}~\bibnamefont{Valenti}},
  \bibinfo{author}{\bibfnamefont{H.}~\bibnamefont{Rosner}}, \bibnamefont{and}
  \bibinfo{author}{\bibfnamefont{C.}~\bibnamefont{Gros}},
  \bibinfo{journal}{Cond-mat/0312156v2}  (\bibinfo{year}{2004}).

\bibitem[{\citenamefont{Imai and Chou}(2003)}]{imait2003}
\bibinfo{author}{\bibfnamefont{T.}~\bibnamefont{Imai}} \bibnamefont{and}
  \bibinfo{author}{\bibfnamefont{F.~C.} \bibnamefont{Chou}},
  \bibinfo{journal}{Cond-mat/0301425}  (\bibinfo{year}{2003}).

\bibitem[{\citenamefont{de~Wolff et~al.}(1981)\citenamefont{de~Wolff, Janssen,
  and Janner}}]{dewolff1981}
\bibinfo{author}{\bibfnamefont{P.~M.} \bibnamefont{de~Wolff}},
  \bibinfo{author}{\bibfnamefont{T.}~\bibnamefont{Janssen}}, \bibnamefont{and}
  \bibinfo{author}{\bibfnamefont{A.}~\bibnamefont{Janner}},
  \bibinfo{journal}{Acta Crystallogr. A} \textbf{\bibinfo{volume}{37}},
  \bibinfo{pages}{625} (\bibinfo{year}{1981}).

\bibitem[{\citenamefont{van Smaalen}(1995)}]{vansmaalen1995}
\bibinfo{author}{\bibfnamefont{S.}~\bibnamefont{van Smaalen}},
  \bibinfo{journal}{Crystallogr. Rev.} \textbf{\bibinfo{volume}{4}},
  \bibinfo{pages}{79} (\bibinfo{year}{1995}).

\bibitem[{\citenamefont{Caimi et~al.}(2004{\natexlab{b}})\citenamefont{Caimi,
  Degiorgi, Lemmens, and Chou}}]{caimig2004b}
\bibinfo{author}{\bibfnamefont{G.}~\bibnamefont{Caimi}},
  \bibinfo{author}{\bibfnamefont{L.}~\bibnamefont{Degiorgi}},
  \bibinfo{author}{\bibfnamefont{P.}~\bibnamefont{Lemmens}}, \bibnamefont{and}
  \bibinfo{author}{\bibfnamefont{F.~C.} \bibnamefont{Chou}},
  \bibinfo{journal}{Cond-mat/0404502}  (\bibinfo{year}{2004}{\natexlab{b}}).

\end{thebibliography}

\end{document}